%% file: main.tex
\documentclass[sigconf,authorversion=true]{acmart}
\usepackage[utf8]{inputenc}
\usepackage[T1]{fontenc}
\usepackage{amssymb}
\usepackage{pifont}
\newcommand{\cmark}{\ding{51}}%
\newcommand{\xmark}{\ding{55}}%

\usepackage{setspace}
\usepackage{booktabs} 

\usepackage[caption=false,font=footnotesize]{subfig}
\usepackage{fancyhdr}
\usepackage{mdwlist}
\usepackage{multirow}
\usepackage{amsmath}        

\graphicspath{{incl/}}

\makeatletter
\def\blfootnote{\xdef\@thefnmark{}\@footnotetext}
\makeatother

\def\smallerspacecaption{\vspace{-3mm}}

\definecolor{gray}{gray}{0.9}

\newcommand{\drop}[1]{}

\begin{document}

\setlength{\textfloatsep}{9pt plus 2pt minus 4pt}
\setlength{\floatsep}{6pt plus 2pt minus 2pt}
\setlength{\dbltextfloatsep}{9pt plus 2pt minus 4pt}
\setlength{\dblfloatsep}{6pt plus 2pt minus 2pt}


\input{footer}
\input{classification}

\title{Protect Your Chip Design Intellectual Property: An Overview}

\author{Johann Knechtel,
	Satwik Patnaik,
	and Ozgur Sinanoglu
		}
 \authornote{J.\ Knechtel and S.\ Patnaik contributed equally.
		 This work was supported in part by NYUAD REF under Grant RE218
		 and by NYU/NYUAD CCS.}
 \affiliation{Tandon School of Engineering, New York University, New York, USA\\
		 Division of Engineering, New York University Abu Dhabi, United Arab Emirates}
 \email{{johann, sp4012, ozgursin}@nyu.edu}

\begin{abstract}
\input{abstract}
\end{abstract}

\keywords{
		Logic Locking, Layout Camouflaging, Split Manufacturing}

\maketitle

\section{Introduction}
\label{sec:introduction}
As our modern lives are more and more dependent on ubiquitous information technology, it is critical,
yet highly challenging, to ensure the security and trustworthiness of the underlying integrated circuits (ICs).
For example, researchers have cautioned against powerful attacks on the speculative execution of
processor ICs~\cite{kocher18,lipp18}, or profiled the side-channel leakage of
cryptographic modules~\cite{lerman18}.
Besides such concerns regarding security at runtime, protecting against other threats such as reverse
engineering (RE), intellectual property (IP) piracy, illegal overproduction, or insertion of hardware Trojans (HTs) is another 
challenge.
Note that these threats arise due to the globalized and distributed nature of modern IC supply chains, 
which span
across many parties and countries~\cite{rostami14}.
Over the last decade, a multitude of protection schemes have been proposed (and selectively already implemented in silicon), which can be broadly
classified into
\emph{logic locking},
\emph{layout camouflaging},
and \emph{split manufacturing}.
All these techniques seek
to protect the hardware
from different attackers,
     as summarized in
Table~\ref{tab:protection_comparison}.

\begin{table}[t]
\centering
\footnotesize
\caption{IP Protection Techniques Versus Untrusted Entities
	(\cmark: Protection Offered, \xmark: No Protection Offered)
\label{tab:protection_comparison}
}
\setlength{\tabcolsep}{1.4mm}
\renewcommand{\arraystretch}{1.20}
\label{techniques}
\smallerspacecaption
\begin{tabular}{|*{4}{c|}}
\hline
\textbf{Technique}
& \textbf{FEOL/BEOL Foundry} 
& \textbf{Test Facility} 
& \textbf{End-User} \\
\hline
\hline
Logic Locking
& \cmark/\cmark & \cmark~\,\,(see also~\cite{yasin16_test}) & \cmark   \\ \hline
Layout Camouflaging
& \xmark/\xmark~\,\,(\cmark/\xmark~\cite{patnaik17_Camo_BEOL_ICCAD},
	\cmark/\cmark~\cite{rangarajan18_MESO_arXiv})
& \xmark~\,\,(\cmark~\cite{rangarajan18_MESO_arXiv})
& \cmark   \\ \hline
Split Manufacturing
& \cmark/\xmark~\,\,(\xmark/\cmark~\cite{wang17_FEOL}) & \xmark & \xmark~\,\,(\cmark~\cite{patnaik18_3D_ICCAD,gu2018cost}) \\ \hline
\end{tabular}
\vspace{-1mm}
\end{table}

In this paper, we present an overview of the different schemes while discussing the threat models, recent developments, attack avenues,
limitations, and future directions for research.
This paper also serves as a follow-up to an 
earlier survey~\cite{rajendran2014regaining}.

\section{Logic Locking}
\label{sec:logic_locking}

\subsection{Concept}
Logic locking (LL) protects the IP by inserting dedicated locks
which are operated by a secret key. 
The locks are commonly realized by additional, interposed logic (e.g., XOR/XNOR gates,
AND/OR gates
or look-up tables (LUTs)).
Only after manufacturing (but before deployment), the locked IC is to be activated by loading the secret key 
onto a dedicated, tamper-proof on-chip memory.
Note that the secure realization of tamper-proof memories remains under research and development~\cite{tuyls06,anceau17}.

\subsection{Threat Model}
\label{sec:logic_locking_TM}

\begin{figure*}[tb]
	\centering
	\includegraphics[width=.995\textwidth, trim = {8mm 0mm 0mm 0mm}, clip = true]{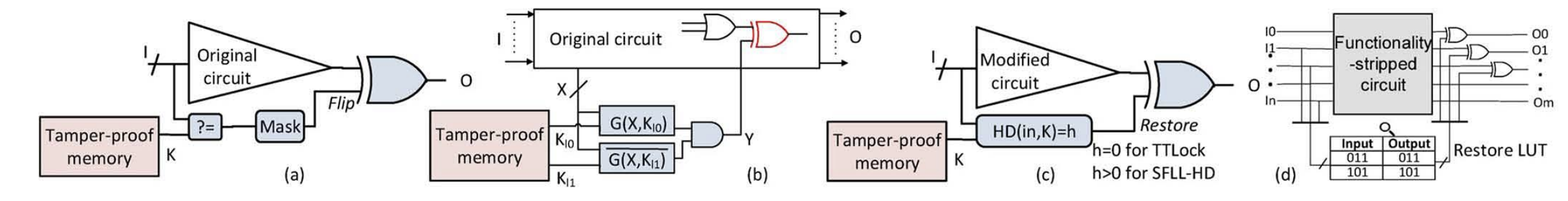} 
    \smallerspacecaption
	\caption{Selected, SAT-resilient locking schemes: (a) \emph{SARLock}, (b) \emph{Anti-SAT}, (c)
		\emph{TTLock}/\emph{SFLL-HD}, and (d) \emph{SFLL-flex}.
	\copyright~2017 IEEE. Reprinted, with permission,
	from~\cite{yasin17_VLSISOC}.
	\label{fig:SAT_resilient_schemes}
	}
\end{figure*}

In general, a threat model describes the attackers' capabilities and the resources at their disposal. 
It also classifies entities 
as trusted or untrusted.
The threat model for LL can be summarized as follows:
\begin{itemize}
\item The design house is considered \emph{trusted},
and so are the designers as well as the electronic design automation (EDA) tools they work with, whereas
the foundry, the test facility, 
and the end-user(s) are all considered \emph{untrusted}.
\item The attackers know the LL scheme which has been applied.
\item The attackers have access to the locked netlist (e.g., by RE). Hence, they can identify the key inputs and the related logic,
but are oblivious to the actual, secret key.
\item The secret key cannot be tampered with, 
as it is programmed in a tamper-proof memory.
\item The attackers are in possession of an already functional chip, e.g., bought from the open market. This chip can act as an ``oracle''
for evaluating input/output patterns.
\end{itemize}

Without knowledge of the secret key, 
LL ensures that:
(i)~the details of the original design cannot be fully recovered;
(ii)~the IC is non-functional, i.e., it produces
incorrect outputs;
and (iii)~targeted insertion of HTs is difficult---an attacker, in absence of the recovered design, cannot readily locate the
	appropriate places to insert HTs.

\subsection{Logic Locking Schemes and Attacks}

Early research proposed
random locking (\emph{RLL})~\cite{roy10},
fault-analysis-based locking (\emph{FLL})~\cite{JV-Tcomp-2013}, and strong interference-based locking
(\emph{SLL})~\cite{JV_DAC_2012}---protecting against brute-force and other simple attacks.
These techniques identify suitable but selected locations for embedding the key;
multiple attacks have undermined them~\cite{JV_DAC_2012,plaza15,li19piercing}.

In 2015, Subramanyan \emph{et al.}~\cite{subramanyan15}
challenged the security of all then-known
LL schemes.
Their attack leverages Boolean satisfiability (SAT) to compute 
so-called \emph{discriminating input patterns (DIPs)}.
A DIP generates different outputs for the same input across two (or more) different keys, indicating that
   at least one of the keys is incorrect.
The attack step-wise evaluates different DIPs until all incorrect keys have been pruned.
The attack experiences its worst-case scenario when it can eliminate only one incorrect key per DIP; here,
$2^k-1$ DIPs are required to resolve $k$ key bits.
In general, the SAT attack resilience of any locking scheme can be represented by the number of DIPs required to 
decipher the correct key~\cite{xie16_SAT}.

In 2016,
\emph{SARLock}~\cite{yasin16_SARLock} and \emph{Anti-SAT}~\cite{xie16_SAT} were put forward as defense schemes
against the SAT-based attack~\cite{subramanyan15}.
\emph{SARLock} (Fig.~\ref{fig:SAT_resilient_schemes}(a)) employs controlled corruption of the output, across all incorrect keys,
for exactly one input pattern.
\emph{SARLock} can also be integrated with other high-corruptibility schemes (e.g., \emph{FLL} or \emph{SLL}) to provide a two-layer defense.
In \emph{Anti-SAT}~\cite{xie16_SAT}, two complementary logic blocks, embedded with the key gates, 
converge at an AND gate (Fig.~\ref{fig:SAT_resilient_schemes}(b)). 
The output of this AND gate is always `0' for the correct key; for the incorrect key, it may be `1' or `0', depending on the inputs.
This AND gate then feeds an additional XOR gate which is interposed into the original design, thereby possibly inducing incorrect outputs
for incorrect keys.
Both schemes utilize the concept of \emph{one-point functions} and enforce low output corruptibility 
to obtain resilience against the SAT-based attack. 

The two-layer defense of \emph{SARLock} was \emph{approximately} circumvented by
\emph{AppSAT}~\cite{shamsi17,shamsi18_TIFS} and \emph{Double DIP}~\cite{shen17}. 
In both the attacks, the combination 
of a low-corruption part (resilient to SAT attacks) 
and a high-corruption part (prone to SAT attacks) 
is reduced to the low-corruption part (e.g., \emph{SARLock + SLL} to \emph{SARLock}).
Moreover, \emph{Double DIP}~\cite{shen17} can eliminate at least two incorrect keys in each iteration, thereby increasing the attack
efficiency.
For \emph{Anti-SAT}, the two complementary blocks at its heart exhibit significant signal skews, rendering them
distinguishable from other logic,
which is exploited by Yasin \emph{et al.}
in the \emph{signal probability skew (SPS)} attack~\cite{yasin_2017_sps}.
Moreover, both \emph{SARLock} and \emph{Anti-SAT} are
vulnerable to the \emph{bypass attack}~\cite{xu17_BDD}.
This attack picks some key randomly
and determines the inputs that provide incorrect outputs for this chosen key.
Then, additional logic is constructed around the \emph{Anti-SAT}/\emph{SARLock} blocks to recover the overall circuit from these incorrect
outputs.

To summarize, although \emph{SARLock} and \emph{Anti-SAT} demonstrate superior resilience against the seminal SAT
attack~\cite{subramanyan15},
they remain vulnerable to other variants of 
SAT attacks (e.g., \emph{AppSAT}, \emph{Double DIP}) as well as structural attacks 
(e.g., \emph{SPS}, \emph{bypass attack}).
Also note that both schemes keep the to-be-protected IP largely as is, thereby 
opening the doors to \emph{removal attacks}~\cite{yasin17_CCS,yasin17_TETC}.

\subsection{Advanced Logic Locking Schemes}

In \emph{TTLock},
the original logic is modified for exactly one input pattern~\cite{yasin_glsvlsi_2017}.
The output for this protected pattern is restored using a comparator block, as illustrated in Fig.~\ref{fig:SAT_resilient_schemes}(c). 
Even if an attacker succeeds to remove the comparator block, 
she/he obtains a design different from the original one (albeit for only one input pattern).

Following on the heels of \emph{TTLock}, Yasin \emph{et al.}~\cite{yasin17_CCS} proposed \emph{stripped functionality logic locking (SFLL)}.
\emph{SFLL} is resilient against most current attacks, and it enables to trade-off between resilience against SAT attacks
and removal attacks~\cite{yasin17_CCS}. 
It is based on the notion of ``strip and restore,'' where some part of the original
design is removed and the intended functionality is concealed.
The authors also implemented a chip demonstrator in GLOBALFOUNDRIES 65nm technology.
SFLL has three variants, \emph{SFLL-HD}~\cite{yasin17_CCS}, \emph{SFLL-flex}~\cite{yasin17_CCS}, and 
\emph{SFLL-fault}~\cite{sengupta2018atpg,sengupta2018customized}, which we all discuss briefly below.

\emph{SFLL-HD} is a generalized version of \emph{TTLock} which allows the designer to protect a larger number of selected input
patterns.
More specifically, SFLL-HD$^h$ protects $\binom{k}{h}$ input cubes which are Hamming distance (HD) $h$ away from the $k$-bit secret
key.\footnote{Input cubes are partially-specified input patterns; some input bits are set (to `0' or `1')
while others are set to \emph{don't care} (`X').
An n-bit input cube with $k$ set bits (or care bits) encompasses $2^{n-k}$ input patterns~\cite{yasin17_CCS}.
}
The values for $k$ and $h$ dictate the trade-off between SAT attack resilience and removal attack resilience.
It should also be noted that \emph{SFLL-HD}
protects a restricted set of input cubes, 
which are all underpinned by one secret key.
{\em SFLL-flex$^{c \times k}$}, in contrast, allows to protect any $c$ selected input cubes,
each with $k$ specified bits.
Here, the protected patterns
are typically represented using a small set of input cubes, which are then stored in
an on-chip LUT (Fig.~\ref{fig:SAT_resilient_schemes}(d)).
In \emph{SFLL-fault},
fault injection-based heuristics are 
leveraged to identify and protect multiple patterns, and to reduce area cost at the same time.

Both \emph{SFLL-HD} and {\em SFLL-flex$^{c \times k}$} utilize AND-trees which leave structural hints for an opportune attacker.  Such an
attack has been demonstrated recently by Sirone \emph{et al.}~\cite{sirone18}; the authors deciphered the key successfully (even
		without oracle access).
At the time of writing,
no attacks have been demonstrated on \emph{SFLL-fault} yet.

Shamsi \emph{et al.}~\cite{shamsi2018cross} presented a layout-centric LL scheme,
based on routing cross-bars comprising obfuscated and configurable vias.

Finally, the notion of \emph{cyclic locking} has been proposed in~\cite{shamsi17_GLSVLSI} and extended in~\cite{roshanisefat18}.
The idea is to create supposedly unresolvable locking instances by introducing feedback cycles.
However, tailored SAT formulations have challenged such locking schemes~\cite{zhou17_CycSAT,shen19}.

\subsection{Future Directions for Logic Locking}

Recent works have proposed \emph{parametric locking}~\cite{yasin_glsvlsi_2017,xie2017delay,zaman2018towards,chakraborty2018gpu}; the
essence is to lock design parameters and profiles.
For example, in~\cite{xie2017delay}, the key not only protects the
functionality of the design but also its timing profile.  A functionally-correct but timing-incorrect key will result in timing violations,
thereby leading to circuit malfunctions.
A \emph{timing-based SAT} attack, presented in~\cite{chakraborty2018timingsat}, circumvented the timing locking approach
in~\cite{xie2017delay}.
Therefore, further research into
\emph{parametric locking} is required.
Finally, \emph{mixed-signal locking} has been advocated recently as well, e.g., in~\cite{jayasankaran2018towards,leonhard19mixlock}.

\section{Layout Camouflaging}
\label{sec:camouflaging}

\subsection{Concept}
\begin{figure}[tb]
\centering
\includegraphics[width=.99\columnwidth]{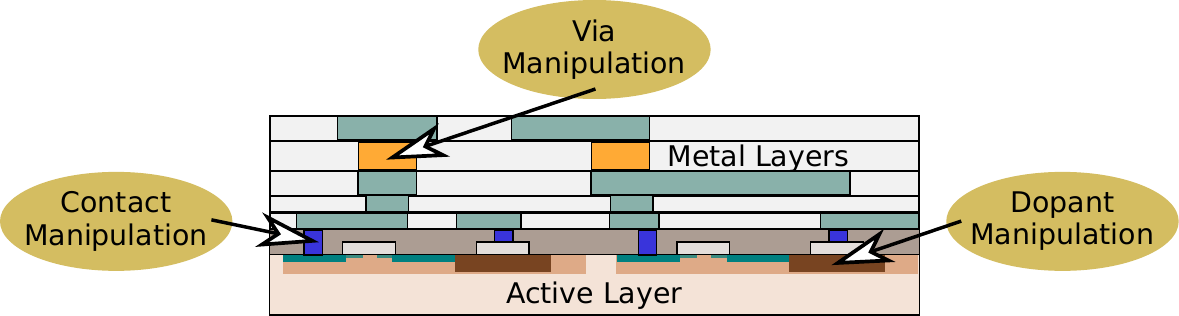}
\smallerspacecaption
\caption{Device-level concepts for camouflaging.
\label{fig:LC}
}
\end{figure}

The objective for layout camouflaging (LC) is to mitigate RE attacks, i.e., reverse engineering of the chip IP (conducted by
malicious end-users). LC seeks to alter the appearance of a chip in order to cloak the chip IP.
That is, LC obfuscates the design information at the device level
(Fig.~\ref{fig:LC}).
Obfuscation can also be conducted at the logic and system level (e.g., obfuscating the finite state
		machine~\cite{lao15}); such techniques are orthogonal to LC.
See also~\cite{vijayakumar16} for an overview.

\subsection{Threat Model}
The threat model for LC is summarized as follows:
\begin{itemize}
\item The design house and foundry are trusted, the test facility is either trusted or untrusted, and the end-user is untrusted.
\item The adversary holds one or multiple functional chip copies, and is armed with more or less sophisticated equipment and know-how to
conduct RE.  The resilience of any LC scheme ultimately depends on the latter.
\item The adversary is aware of the LC scheme,
and she/he can identify the camouflaged gates, infer 
all the possible functions implemented by the camouflaged cell, but cannot readily infer the actual functionality.
\end{itemize}

\subsection{Layout Camouflaging Schemes and Attacks}
Similar to LL, early studies focused on the selection of gates to camouflage (and the design of camouflaged cells).
In their seminal work, Rajendran \emph{et al.}~\cite{rajendran13_camouflage} proposed
a camouflaged NAND-NOR-XOR cell.
The authors also proposed clique-based selection for LC, based on their own finding that
a random selection of gates to camouflage can be resolved by sensitization-based attacks~\cite{rajendran13_camouflage}.
Massad \emph{et al.}~\cite{massad15} and Yu \emph{et al.}~\cite{yu17} formulated independently SAT-based attacks (with oracle access) which 
challenged the security of~\cite{rajendran13_camouflage} nevertheless.\footnote{The essence of these attacks is similar
to~\cite{subramanyan15} and omitted here for brevity; interested readers are also referred to~\cite{massad15,yu17}.}
These attacks could readily circumvent small-scale LC for various benchmarks with up to 256 gates being camouflaged.

A parallel SAT attack providing an average speedup of 3.6$\times$ over prior attacks was presented by Wang \emph{et
	al.}~\cite{wang18_integration}.
Keshavarz \emph{et al.}~\cite{KeshavarzHOST2018} proposed a SAT-based formulation augmented by 
probing and fault injection capabilities, where
the authors were able to RE an \emph{S-Box}. Still, it
remains to be seen whether the attack can tackle larger designs.
In \cite{yasin17_TIFS}, Yasin \emph{et al.} 
demonstrated how an untrusted test facility can circumvent the security promise of LC,
even without access to an oracle.
The authors deciphered LC as in~\cite{rajendran13_camouflage} successfully by analyzing the
test patterns provided by the design house.
To the best of our knowledge, none of the LC schemes proposed thus far 
have been able to mitigate this kind of attack,
except for the \emph{dynamic camouflaging} scheme discussed in~\cite{rangarajan18_MESO_arXiv}.

Many existing LC schemes, e.g., \cite{rajendran13_camouflage,nirmala16,collantes16,wang16_MUX}, exhibit a significant
cost with respect to power, performance, and area (PPA).
For example, the NAND-NOR-XOR gate proposed in~\cite{rajendran13_camouflage} incurs 5.5$\times$
power, 1.6$\times$ delay, and 4$\times$ area cost
when compared to a regular 2-input NAND gate. 
A detailed investigation of PPA cost for various schemes is given in~\cite{patnaik17_Camo_BEOL_ICCAD}. 
Most LC schemes also 
require modifications for the front-end-of-line (FEOL) manufacturing
process, which can incur financial cost on top of PPA overheads.
Therefore, LC is applied rather selectively, to limit PPA cost and the
impact on FEOL processing. 
As indicated above, however, the selective application of LC schemes can compromise their security, especially in the light of
\emph{oracle-guided SAT} attacks such as~\cite{massad15,yu17,wang18_integration}.

The notion of \emph{provably secure camouflaging} 
was put forward in~\cite{yasin16_CamoPerturb,li16_camouflaging}.
\emph{CamoPerturb}~\cite{yasin16_CamoPerturb} 
seeks to minimally perturb the functionality of the design 
by either removing or adding one \emph{minterm} (i.e., the product term of all variables).
A separate block, called \emph{CamoFix}, is then added to restore
the minterm; \emph{CamoFix} is built up using
camouflaged INV/BUF cells.
Inspired by LL, Li \emph{et al.}~\cite{li16_camouflaging}
leverage AND-trees as well as OR-trees for LC.
Depending on the desired security level, tree structures inherently present in the design are leveraged, or additional trees are inserted.
Then, the inputs of the trees are camouflaged using dopant-obfuscated cells.

Both techniques~\cite{yasin16_CamoPerturb,li16_camouflaging} have been shown to exhibit vulnerabilities:
\cite{li16_camouflaging}
was circumvented by a so-called \emph{sensitization-guided SAT 
attack (SGS)}~\cite{yasin17_TETC}, while Jiang \emph{et al.}~\cite{jiang2018efficient} circumvented \emph{CamoPerturb} using
\emph{sensitization} and \emph{implication} principles leveraged from automated test pattern generation (ATPG).
In general, these schemes are also vulnerable to \emph{approximate} attacks outlined in~\cite{shamsi17,shamsi18_TIFS,shen17}.
A follow-up work to~\cite{li16_camouflaging} is presented in~\cite{li17_TCAD}, where the authors discuss how structural attacks like
\emph{SPS}~\cite{yasin_2017_sps}
can be rendered ineffective when the trees are obfuscated both structurally and functionally.

Besides the various analytical attacks, RE may also compromise LC schemes directly.
For example, ambiguous gates~\cite{rajendran13_camouflage,cocchi14} or secretly configured MUXes~\cite{wang16_MUX} rely on dummy contacts
and/or dummy channels, which will induce different charge accumulations at runtime.  Courbon \emph{et al.}~\cite{courbon16} leveraged
scanning electron microscopy in the passive voltage contrast mode (SEM PVC) for measurement of charge accumulations, whereupon they
succeeded in reading out a secured memory.
Furthermore, monitoring the photon emission at runtime, as for example
proposed by Lohrke \emph{et al.}~\cite{lohrke16}, can presumably also help to uncover LC.

\subsection{Advanced Layout Camouflaging Schemes}
Threshold voltage-based camouflaging (TVC) has gained significant traction recently.
The essence of TVC is a selective manipulation of dopants at the transistor level,
to create cells which are identical structurally but operate with different functionality.
Nirmala \emph{et al.}~\cite{nirmala16} proposed TVC cells which can operate as NAND, NOR, OR, AND, XOR, or XNOR.
Erbagci \emph{et al.}~\cite{erbagci16} proposed TVC cells operating as XOR or XNOR, based on the selective use of high- and low-threshold transistors.
Collantes \emph{et al.}~\cite{collantes16} adopted \emph{domino logic} to implement their TVCs.
Recently, Iyengar \emph{et al.}~\cite{iyengar2018threshold}
demonstrated two flavors of TVC in STMicroelectronics 65nm technology.
In principle, TVC schemes offer better resilience than other LC schemes as
regular etching and optical-imaging techniques are ineffective.
Still, TVC may be revealed eventually, e.g., by leveraging SEM PVC~\cite{sugawara14}.

Another interesting avenue is the camouflaging of the back-end-of-line (BEOL), i.e., the
interconnects~\cite{chen15,patnaik17_Camo_BEOL_ICCAD, shamsi2018cross,jang2018threshold}.
Chen \emph{et al.}~\cite{chen15, chen18_interconnects} explored the use of real vias (magnesium, Mg) along with dummy vias
(magnesium oxide, MgO). The authors (and others, e.g., \cite{hwang12_transient_electronics})
have shown that Mg can oxidize quickly into MgO, thereby hindering an identification
by an RE attacker.
Recently, Patnaik \emph{et al.}~\cite{patnaik17_Camo_BEOL_ICCAD} extended the concept of BEOL camouflaging in conjunction
with \emph{split manufacturing} (Sec.~\ref{sec:split_manufacturing}),
to protect against an untrusted FEOL foundry, which was a first for LC. 
Patnaik \emph{et al.}
developed customized cells and design stages for BEOL camouflaging, whereupon they succeeded to
demonstrate full-chip camouflaging at lower PPA cost than prior works.
Their study also explored how large-scale (BEOL) camouflaging can thwart SAT-based attacks, ``simply'' by inducing overly large and
complex SAT instances.

\subsection{Future Directions for Layout Camouflaging}
Most schemes discussed so far cannot be configured post-fabrication, i.e., they implement static camouflaging.
In contrast, Akkaya \emph{et al.}~\cite{akkaya18} demonstrated a reconfigurable
LC scheme
which leverages hot-carrier injection.
The authors succeeded to fabricate a prototype in 65nm technology;
however, they
report significant PPA cost
(e.g., in comparison to regular NAND gates, they report 9.2$\times$, 6.6$\times$, and 7.3$\times$ for power, performance, and area,
		respectively).

Zhang \emph{et al.}~\cite{zhang18_TimingCamo}
introduced \emph{timing-based LC}, based on wave-pipelining and
false paths. However, this scheme was circumvented in~\cite{li2018timingsat}.
Besides, emerging devices are gaining traction as well in the context of
LC~\cite{bi16_JETC,parveen17,patnaik18_GSHE_DATE,rangarajan18_MESO_arXiv}.
For example, Rangarajan \emph{et al.}~\cite{rangarajan18_MESO_arXiv} explore magneto-electric spin-orbit (MESO) devices for reconfigurable LC.

In short, future
work should seek to make LC more RE-resilient, leverage new devices and circuit principles,
reduce dependencies on foundries, and yet enable low overheads.

\section{Split Manufacturing}
\label{sec:split_manufacturing}

\subsection{Concept}
\begin{figure}[tb]
\centering
\includegraphics[width=.86\columnwidth]{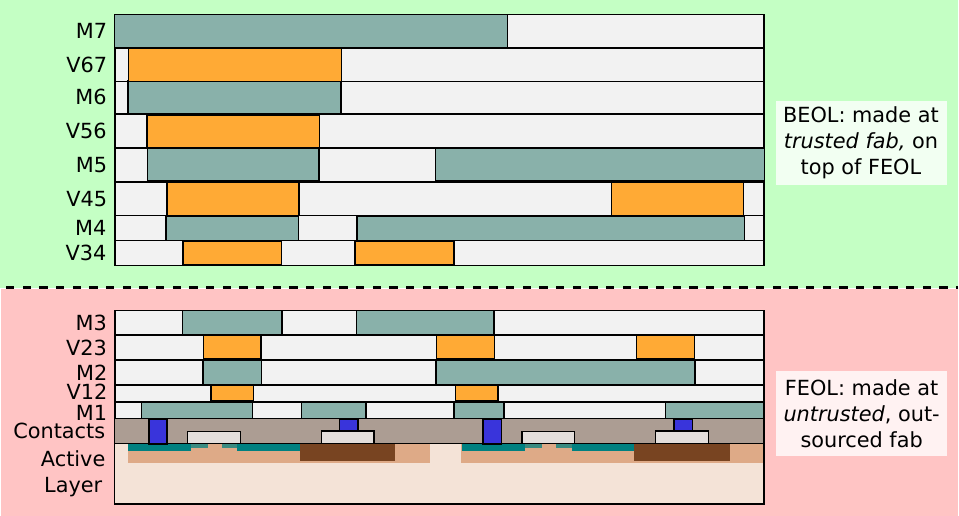}
\smallerspacecaption
\caption{Classical split manufacturing, i.e., 
the separation into FEOL and BEOL parts.
	\copyright~2018 IEEE. Reprinted, with permission,
	from~\cite{patnaik18_SM_ASPDAC}.
\label{fig:SM_concept}
}
\vspace{-1.5mm}
\end{figure}

Split manufacturing (SM) protects the design 
IP from untrustworthy foundries during manufacturing time~\cite{mccants11}.
The idea is to split up the manufacturing flow, typically
into the FEOL and BEOL process steps (Fig.~\ref{fig:SM_concept}).
Considered individually, the physical layout becomes a ``sea of largely unconnected gates'' for the FEOL foundry, whereas it
becomes system-level wiring without any gate-level information for the BEOL foundry.
Such splitting into FEOL and BEOL is practical for multiple reasons: (i)~outsourcing the FEOL is desired, as it requires some
high-end and costly facilities, (ii)~BEOL fabrication on top of the FEOL is significantly less complex than FEOL fabrication itself,
(iii)~the sole difference for the supply chain is the preparation and shipping of FEOL wafers to
the BEOL facility.

\subsection{Threat Model}
The basic, most common threat model is summarized as follows:
\begin{itemize}
\item The design house and end-user are trustworthy, while the FEOL foundry is deemed untrustworthy.
SM necessitates a trusted BEOL foundry, with assembly and testing 
facilities typically also considered as trustworthy.
\item With the design house and end-user being trusted,
the adversary cannot obtain a chip copy from those entities.
Besides, the chip has typically not been manufactured before; the chip is then unavailable altogether for RE attacks.
\item The objective for the adversary is to infer the missing BEOL connections
from the given but incomplete FEOL layout.
Towards this end, she/he (i)~is aware of the underlying protection scheme, if any, and
(ii)~has access to the utilized EDA tools, libraries, and other supporting information.
\end{itemize}

An ``inverted model'' model was explored in~\cite{wang17_FEOL}, where the BEOL facility is untrustworthy and the FEOL fab is trustworthy.
Since fabricating the FEOL is more costly than the BEOL, the practical relevance of this model remains questionable.

Another variation of the threat model was explored by Chen and Vemuri~\cite{chen18_SAT}.
The authors assume that a working chip is available which is then used as an oracle for a SAT-based formulation to recover the missing
BEOL connections. 
While it is not explicitly stated in~\cite{chen18_SAT}, we presume that the authors seek to recover the
gate-level details of some design whose functionality 
is otherwise already available/known.
For an attacker, doing so can be relevant, e.g., for inserting HTs during re-implementation of some existing design, or to obtain the IP
without RE of the available chip copy.

Imeson \emph{et al.}~\cite{ImesonSEC13} further proposed a ``strong model'' in the context of HTs.
Here, the attacker already holds the 
netlist and is interested in inserting HTs into appropriate locations.
This work~\cite{ImesonSEC13}, also known as \emph{k-security}, has been further extended in~\cite{li18}.

\subsection{Split Manufacturing Schemes and Attacks}
\label{proximity_attacks}

A first attack on SM was proposed by Rajendran \emph{et al.}~\cite{rajendran13_split}.
The notion of this so-called \emph{proximity attack} is as follows: although the layout is split into FEOL and BEOL, it is still
designed holistically (at least when using regular EDA tools);
therefore, various hints on the BEOL can remain in the FEOL.
Rajendran \emph{et al.}~\cite{rajendran13_split} infer from the proximity of cells
which is readily observable in the FEOL,
whether they have to be connected in the BEOL.
While that attack shows a good accuracy for small designs, the same is not true for larger designs.
Wang \emph{et al.}~\cite{wang16_sm}
extended this attack, by taking into account a multitude of FEOL-level hints: (i)~physical proximity of gates, (ii)~avoidance
of combinatorial loops (which are rare in practice), (iii)~timing and
load constraints, and (iv)~orientation of ``dangling wires'' (i.e., the wires remaining unconnected in the top-most FEOL layer).
Maga\~{n}a \emph{et al.}~\cite{magana16} proposed various routing-based attack techniques,
and they conclude that such attacks
are more effective than solely placement-centric attacks.
Recently, Zhang~\emph{et al.}~\cite{zhang2018analysis}
and Li~\emph{et al.}~\cite{li19_SM_ML_DAC}
leveraged machine learning
as an attack framework.
However, neither attack~\cite{zhang2018analysis,magana16,li19_SM_ML_DAC} recovers the actual netlist; rather, they
provide
most probable BEOL connections.

Various techniques have been proposed to safeguard FEOL layouts against proximity attacks, e.g.,
	\cite{rajendran13_split,vaidyanathan14,wang16_sm,sengupta17_SM_ICCAD,wang17,magana16,feng17,patnaik18_SM_ASPDAC,patnaik18_SM_DAC,chen2018improving}.
They can be categorized into (i)~placement-centric, (ii) routing-centric, and (iii) both placement- and routing-centric defenses.

Among others, Wang \emph{et al.}~\cite{wang16_sm} and Sengupta \emph{et al.}~\cite{sengupta17_SM_ICCAD} propose placement perturbation.
Layout randomization is most secure, especially when splitting at the first metal layer, as shown by Sengupta
\emph{et al.}~\cite{sengupta17_SM_ICCAD}.
However, this technique has limited scalability and significant layout cost for larger designs.
In general, placement-centric works caution 
that splitting at higher metal layers---which helps to limit financial cost and practical hurdles for
SM~\cite{xiao15,patnaik18_SM_ASPDAC}---can undermine their resilience.
That is because any placement perturbation is eventually offset by routing at higher layers.

Routing-centric schemes as those in~\cite{rajendran13_split,wang17,magana16,feng17,patnaik18_SM_ASPDAC}
resolve proximity and other hints at the FEOL routing.
Rajendran \emph{et al.}~\cite{rajendran13_split} proposed to swap pins of IP modules and to re-route those nets,
thereby
obfuscating
the design hierarchy.
As these swaps cover only part of the interconnects, this scheme cannot protect against gate-level IP
piracy.
In fact, 87\% of the connections could be correctly recovered on the \emph{ISCAS-85} benchmarks~\cite{rajendran13_split}.
In general, routing-centric schemes are
subject to routing resources and PPA budgets, which can ease proximity attacks.
For example, \cite{wang17,feng17} consider short routing detours, and \cite{magana16} consider few routing blockages.

\subsection{Advanced Split Manufacturing Schemes}
\label{protection_schemes_IP_SM}

Patnaik \emph{et al.}~\cite{patnaik18_SM_ASPDAC} proposed various heuristics as well as
custom cells for lifting wires to the BEOL in a concerted
manner. The authors demonstrated a superior resilience; the state-of-the-art attack~\cite{wang16_sm} could not infer any of the protected
connections correctly.
Later on, Patnaik \emph{et al.}~\cite{patnaik18_SM_DAC} proposed randomization at the netlist level, which is carried through the EDA flow,
thereby resulting in an erroneous and misleading FEOL layout.
The original design is only restored at the BEOL, using customized routing cells.
This work is one of the first to address holistic protection 
of both placement and routing.
The authors also demonstrated superior resilience.

Inspired by LL, Sengupta \emph{et al.}~\cite{sengupta19_LL_SM_DATE} realize IP protection at manufacturing time
by locking the FEOL and subsequent unlocking of the BEOL. The authors also formalize the problem of SM.

As mentioned before, Imeson \emph{et al.}~\cite{ImesonSEC13} formulated the notion 
of \emph{k-security}
to prevent targeted insertion of HTs.
The idea is to create \emph{k} isomorphic structures in the FEOL by guided lifting of wires to the BEOL.  Now, an attacker cannot uniquely
map these \emph{k} structures to some specific target in the already-known design; she/he has to  either randomly guess (with a probability of
$1/k$) or insert multiple HTs.
Li \emph{et al.}~\cite{li18} extended \emph{k-security} in various ways. Most notably, they
leverage additional gates and wires to be able to
elevate the security levels beyond those achieved in~\cite{ImesonSEC13}. 
Recently, Xu \emph{et al.}~\cite{xu19} questioned the theoretical security of \emph{k-security} by pattern matching attacks conducted on the
layout level.

Vaidyanathan \emph{et al.}~\cite{vaidyanathan14_BEOL} advocate testing of the untrusted FEOL against HT insertion,
using BEOL stacks dedicated for testability.

Finally, Xiao \emph{et al.}~\cite{xiao15}
propose the notion of \emph{obfuscated built-in self-authentication
(OBISA)} to hinder IP piracy and HT insertion.

\subsection{Future Directions for Split Manufacturing}

While advanced attacks such as~\cite{zhang2018analysis,li19_SM_ML_DAC} are on the rise, SM becomes inherently more resilient
for larger, industrial designs. In fact, none of the existing attack works
succeeded yet in fully recovering all missing BEOL connections for larger
designs.
Still, the crux for SM---to resolve hints from the FEOL---remains. Thus, schemes 
which further reduce the dependency on EDA tools (and cost) are required.

Although \cite{sengupta19_LL_SM_DATE} explores the formalism of SM,
a notion of \emph{provably secure SM} remains an open problem.
Finally, ``entering the next dimension of SM,'' by leveraging the up-and-coming techniques for 3D integration, has been initiated 
in~\cite{ImesonSEC13,patnaik18_3D_ICCAD,gu2018cost,valamehr13,knechtel17_TSLDM}. Further research towards this end seems promising as well.

\section{Summary}
\label{sec:summary}

A multitude of techniques have been proposed to protect your chip design from attacks such as 
illegal overproduction, IP piracy, and insertion of Trojans.
We presented an overview on logic locking, layout camouflaging, and split manufacturing---the three main categories for IP
protection.
We
also outlined shortcomings, attack avenues, and
promising directions for future research.

\bibliographystyle{IEEEtran}
\newcommand{\BIBdecl}{\setlength{\itemsep}{0.018pt}}
\input{main.bbl}


\end{document}

%% file: footer.tex
\copyrightyear{2019}
\acmYear{2019}
\setcopyright{acmcopyright}
\acmConference[COINS]{INTERNATIONAL CONFERENCE ON OMNI-LAYER
INTELLIGENT SYSTEMS}{May 5--7, 2019}{Crete, Greece}
\acmBooktitle{INTERNATIONAL CONFERENCE ON OMNI-LAYER
INTELLIGENT SYSTEMS (COINS), May 5--7, 2019, Crete, Greece}
\acmPrice{15.00}
\acmDOI{10.1145/3312614.3312657}
\acmISBN{978-1-4503-6640-3/19/05}

%% file: classification.tex
%

\ccsdesc[500]{Security and privacy~Security in hardware}

%% file: abstract.tex
The increasing cost of integrated circuit (IC) fabrication has driven most companies to ``go fabless'' over time.
The corresponding outsourcing trend gave rise to various attack vectors, e.g., 
illegal overproduction of ICs, piracy of the design intellectual property (IP), or insertion of hardware Trojans (HTs).
These attacks are possibly conducted by untrusted entities
residing all over the supply chain, ranging from untrusted foundries, test facilities, even to end-users.
To overcome this multitude of threats, various techniques have been proposed over the past decade.
In this paper, we review the landscape of IP protection techniques, which can be classified into
logic locking, layout camouflaging, and split manufacturing. 
We discuss the history of these techniques, followed by state-of-the-art advancements,
relevant limitations, and scope for future work.